\documentclass[reprint,aps,prl,twocolumn,superscriptaddress,nobibnotes]{revtex4-2}

\usepackage{graphicx}
\usepackage[margin=1in]{geometry}
\usepackage{dcolumn}
\usepackage{bm}
\usepackage{braket}
\usepackage{xfrac}
\usepackage[separate-uncertainty = true, multi-part-units=single]{siunitx}
\usepackage{xcolor}
\usepackage{gensymb}
\usepackage{tikz}
\usepackage[percent]{overpic}

\usepackage[utf8]{inputenc}
\usepackage[T1]{fontenc}
\usepackage{lmodern}
\usepackage{etoolbox}

\usepackage{natbib}

\begin{document}

\preprint{APS/123-QED}

\title{Cryogenic nonlinear processes in thin-film lithium niobate}

\author{Tristan Kuttner}
\thanks{These authors contributed equally to this work.}
\affiliation{Optical Nanomaterial Group, Institute for Quantum Electronics, Department of Physics, ETH Zurich, CH-8093 Zurich, Switzerland}

\author{Ulrich Sauter}
\thanks{These authors contributed equally to this work.}
\affiliation{Optical Nanomaterial Group, Institute for Quantum Electronics, Department of Physics, ETH Zurich, CH-8093 Zurich, Switzerland}

\author{Robert J. Chapman}
\thanks{These authors contributed equally to this work.}
\email{rchapman@ethz.ch}
\affiliation{Optical Nanomaterial Group, Institute for Quantum Electronics, Department of Physics, ETH Zurich, CH-8093 Zurich, Switzerland}

\author{Myriam Rihani}
\affiliation{Optical Nanomaterial Group, Institute for Quantum Electronics, Department of Physics, ETH Zurich, CH-8093 Zurich, Switzerland}

\author{Jost Kellner}
\affiliation{Optical Nanomaterial Group, Institute for Quantum Electronics, Department of Physics, ETH Zurich, CH-8093 Zurich, Switzerland}

\author{Alessandra Sabatti}
\affiliation{Optical Nanomaterial Group, Institute for Quantum Electronics, Department of Physics, ETH Zurich, CH-8093 Zurich, Switzerland}

\author{Giovanni Finco}
\affiliation{Optical Nanomaterial Group, Institute for Quantum Electronics, Department of Physics, ETH Zurich, CH-8093 Zurich, Switzerland}

\author{Andreas Maeder}
\affiliation{Optical Nanomaterial Group, Institute for Quantum Electronics, Department of Physics, ETH Zurich, CH-8093 Zurich, Switzerland}

\author{Rachel Grange}
\affiliation{Optical Nanomaterial Group, Institute for Quantum Electronics, Department of Physics, ETH Zurich, CH-8093 Zurich, Switzerland}

\date{\today}

\begin{abstract}

Photonic integrated circuits operating at cryogenic temperatures are necessary for many quantum technologies such as quantum transduction, integrated single-photon emitters and detectors, as well as deep-space communication and sensing devices. 
Thin-film lithium niobate (TFLN) is an emerging platform that is a strong candidate for fully integrated quantum photonics, offering low loss, fast electro-optic reconfigurability, nonlinear quantum light sources, and the ability to host quantum emitters and single-photon detectors.
To interface TFLN with technologies that require cryogenic operation, like superconducting single-photon detectors, microwave-to-optical transducers, and solid-state quantum emitters, it is important to study its optical and electrical properties from room temperature down to cryogenic temperatures.
Here, we investigate linear and nonlinear photonic devices, including racetrack resonators, Mach-Zehnder modulators and periodically poled waveguides in TFLN using a cryogenic fiber probe station with full temperature control down to \SI{5}{\K}.
We quantify a shift in resonances, a \SI{22}{\%} increase in electro-optic modulator half-wave voltage, a blue shift of \SI{18}{\nm} for Type-0 phase-matching as well as a red shift of \SI{64}{\nm} for Type-II phase-matching as the sample temperature decreases.
Our study of nonlinear processes in a cryogenic environment will contribute towards developing novel devices for inter-platform quantum information processing, secure communication, and enhanced sensing.

\vspace{1.5cm}
\end{abstract}

\maketitle

\subsection*{Introduction}

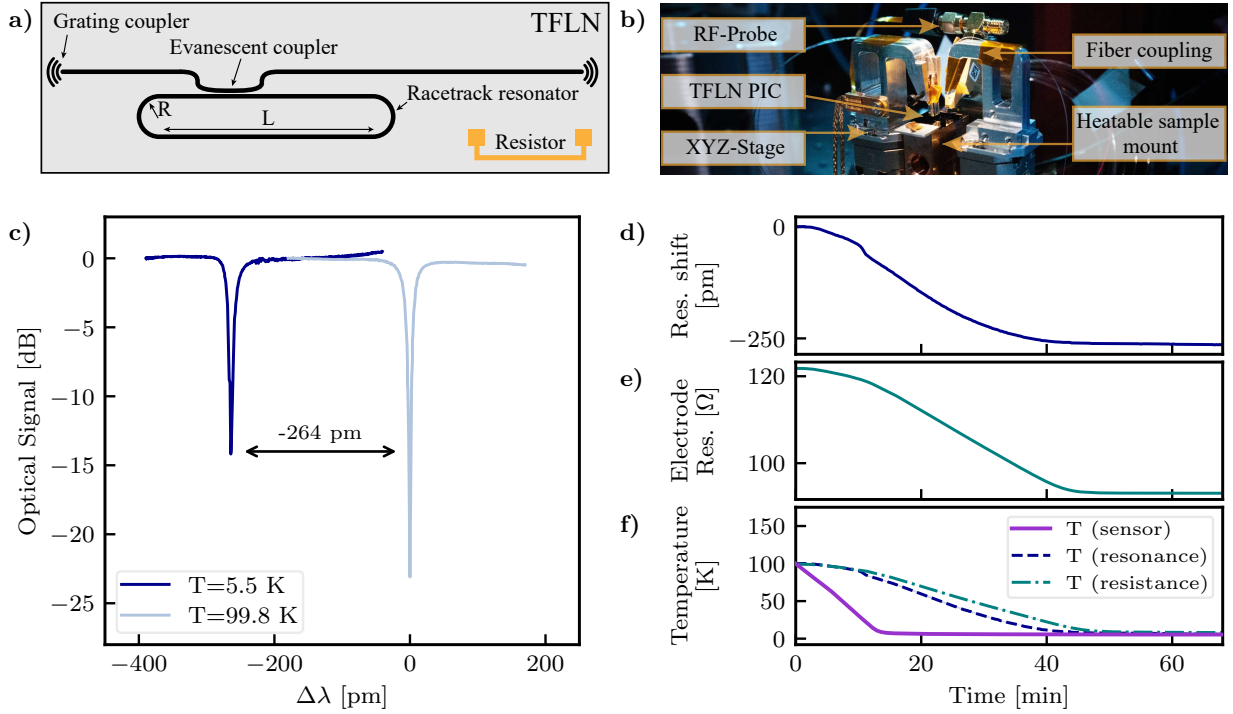
\begin{figure*}[!t]
    \centering
        \includegraphics[width=1.0\linewidth]{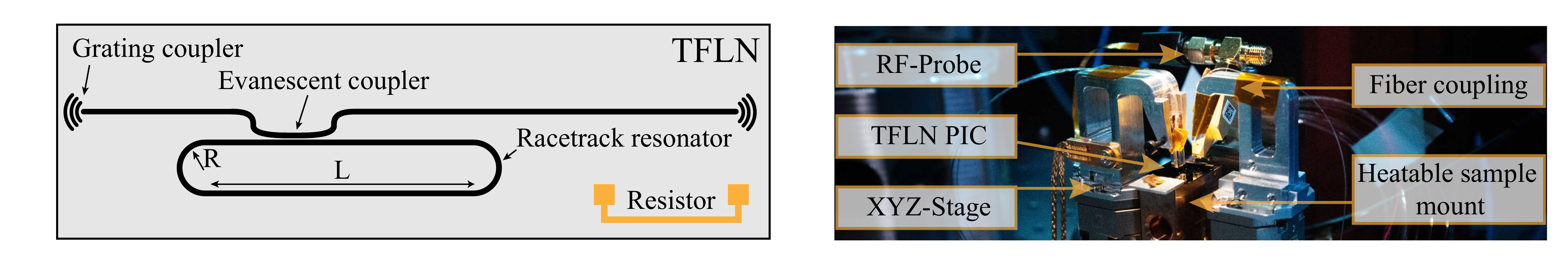}
        \includegraphics[width=1.0\linewidth]{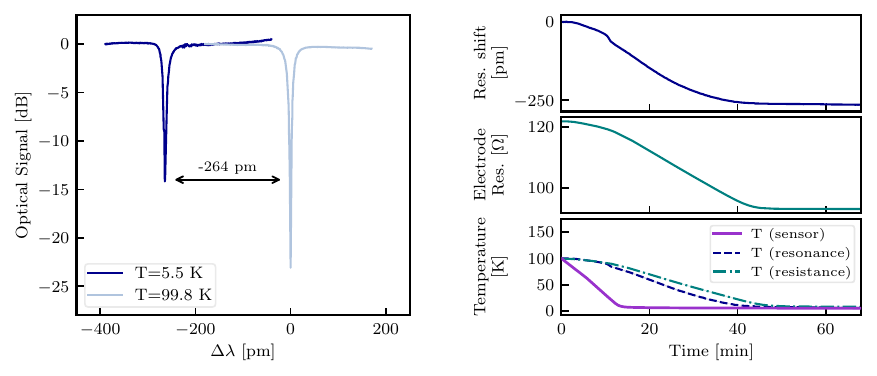}

    \begin{picture}(0,0)
        \put(-230,275){\textbf{a)}}
        \put(0,275){\textbf{b)}}
        \put(-230,195){\textbf{c)}}
        \put(0,195){\textbf{d)}}
        \put(0,140){\textbf{e)}}
        \put(0,85){\textbf{f)}}
    \end{picture}
    \caption{\textbf{TFLN photonic integrated circuit thermalization.}
    (a) Device layout for the thermalization characterization including a racetrack resonator and gold micro-resistor.
    (b) Automated fiber probe station inside the temperature-controlled cryostat.
    (c) The shift of a single resonance from \SI{99.8}{\K} to \SI{5.5}{\K}.
    (d-f) Racetrack resonance wavelength, on-chip micro-resistor resistance, and cryostat temperature sensor reading during the last hour of a cooldown cycle, with inferred temperatures from the resonance shift and resistance change.
    }
    \label{fig:thermal}
\end{figure*}

Photonic integrated circuits have a vast range of applications in classical and quantum technology \cite{shekhar_roadmapping_2024, wang_integrated_2020}, including frequency comb generation \cite{diddams_optical_2020}, optical neural networks \cite{fu_optical_2024}, sources of entangled photons \cite{finco_time-bin_2024}, transduction for quantum networking \cite{holzgrafe_cavity_2020}, and quantum computing \cite{larsen_integrated_2025}.
Many applications of quantum photonic circuits require operation at cryogenic temperatures, such as integration of single-photon emitters \cite{davanco_heterogeneous_2017}, on-chip superconducting nanowire single-photon detectors (SNSPDs) \cite{ferrari_waveguide-integrated_2018}, and transduction between optical and microwave photons \cite{holzgrafe_cavity_2020}.
Although silicon and silicon nitride have been dominant platforms for advanced integrated photonic circuits, they lack a second-order nonlinearity and therefore rely on micro-heaters to apply phase shifts for reconfigurability, which are not compatible with on-chip single-photon emitters and SNSPDs without challenging thermal management \cite{elshaari_-chip_2017, alexander_manufacturable_2025}.
Additionally, a large quantum photonic circuit with hundreds of thermo-optic phase shifters might dissipate high amounts of thermal power, easily exceeding the cooling power of the most powerful commercial cryostats.
This pitfall of thermo-optic reconfigurable devices has motivated research into integrated nanophotonic platforms with electro-optic tunability such as thin-film lithium niobate (TFLN), which can achieve very high bandwidths with little energy dissipation \cite{zhu_integrated_2021}. 
The electro-optic effect and tight mode confinement in TFLN photonics have enabled high-speed modulators with sub-volt signals \cite{wang_integrated_2018}, frequency combs \cite{zhang_broadband_2019} and frequency shearing \cite{zhu_spectral_2022}.
Electro-optic phase shifters dissipate nearly zero thermal energy, making them ideal for cryogenic photonic integrated circuits.

Furthermore, due to its ferroelectricity, the non-centrosymmetric lithium niobate (LN) crystal can be periodically poled to achieve quasi-phase-matching, where periodically poled TFLN photonics has enabled ultra-efficient frequency conversion \cite{wang_ultrahigh-efficiency_2018, lu_periodically_2019}, integrated optical parametric oscillators \cite{lu_ultralow-threshold_2021, kellner_low_2025}, photon-pair generation \cite{zhao_high_2020, finco_time-bin_2024, chapman_-chip_2025, kuttner_heralded_2026, kellner_counter-propagating_2026, cheng_efficient_2025}, squeezed light generation \cite{nehra_few-cycle_2022, stokowski_integrated_2023}, and quantum frequency conversion \cite{wang_quantum_2023}.
Combining on-chip quantum light sources and high-speed electro-optic reconfigurability makes TFLN a promising platform for scaling up integrated quantum photonics \cite{sund_high-speed_2023}.
While the electro-optic effect and quasi-phase-matching have been studied at low temperatures with titanium in-diffused waveguides in bulk LN \cite{thiele_cryogenic_2022, bartnick_cryogenic_2021}, a comprehensive study of cryogenic TFLN photonics is still lacking.
Therefore, it is necessary to investigate the behavior and performance of TFLN photonic integrated circuits at cryogenic temperatures for future applications, including on-chip single-photon detection, integration with other systems such as transducers or quantum emitters, and operation in deep-space environments.

Here, we investigate the properties of x-cut TFLN photonic integrated circuits from room temperature (RT) down to \SI{5}{K}.
We first study the thermalization times of the sample by probing the resonance shift of an on-chip racetrack resonator as well as the resistance of an on-chip gold micro-resistor.
Next, we measure the efficiency of a TFLN electro-optic modulator, where we observe a $\sim$\SI{22}{\%} increase in $V_{\pi}$ at cryogenic temperatures, indicating a decrease of the $r_{33}$ electro-optic coefficient, which is in line with our simulations considering the decrease in $r_{33}$ and $\epsilon_{33}$, as reported for bulk lithium niobate \cite{herzog_electro-optic_2008}.
Furthermore, we investigate nonlinear frequency conversion in periodically poled TFLN waveguides by measuring second harmonic generation (SHG) in a Type-0 process, which yields a blue-shift of the phase-matching wavelength of \SI{18}{nm} going from RT to \SI{5}{K}. 
Finally, we measure Type-II phase-matching, where we perform sum frequency generation (SFG) and observe a red-shift of the phase-matching function by \SI{64}{nm} over the same temperature range.

\subsection*{Setup and Thermalization}

Firstly, we study the thermalization of the TFLN samples during a change of the operation temperature of the cryostat.
We use a \SI{300}{nm} thick LN sample with \SI{230}{nm} etch depth, a \SI{1}{\um} silicon dioxide cladding, and a \SI{500}{\um} thick silicon handle, featuring optical structures with waveguide widths of \SI{1.2}{\um}.
The device layout of the used sample can be seen in Fig. \ref{fig:thermal}(a).
The sample is placed in a cryogenic chamber on a copper mount which is thermally anchored to the cold-head by copper braids.
This mount features a heater and temperature sensor; the cryostat itself is equipped with fiber, DC and RF feedthroughs, as seen in the picture in Fig. \ref{fig:thermal}(b).
The TFLN chips are clamped to the sample mount, thus, we anticipate a thermalization delay between the sensor and the actual chip temperature when controlling the sample stage heater.
It is crucial to take this effect into account during our experiments, since we want to probe as a function of the chip temperature.
Therefore, we characterize the thermalization process with two different approaches: a racetrack resonator with bend radii of \SI{80}{\um} and straight lengths of \SI{1.4}{mm}, which results in a free spectral range of \SI{333}{\pm} at \SI{1545}{\nm}, as well as an integrated gold micro-resistor with a width of \SI{1}{\um}, length of \SI{500}{\um}, and thickness of \SI{100}{\nm}, which acts as an electronic temperature sensor on the thin-film surface.

We measure the transmission of a racetrack resonator as the cryostat cools from 
\SI{99.8}{\K} to \SI{5.5}{K}, where we track a single resonance over this temperature range and measure a shift of \SI{-264}{\pm}, as shown in Fig. \ref{fig:thermal}(c), caused by the temperature-dependent refractive index and thermal contraction of lithium niobate, which is lower than a simulated shift of about \SI{-570}{\pm}.
Additionally, we also record the resistance of the micro-resistor during this cooldown, exhibiting a decrease from \SI{121.8}{\ohm} to \SI{93.1}{\ohm} over the temperature range.
Figure \ref{fig:thermal}(d) and (e) show the resonance shift as well as the resistance change of the micro-resistor over the course of the last hour of a cooldown cycle. 
From these measurements, we can estimate an on-chip temperature by calibrating our simulation models to the measured changes in resonance and resistance of the devices.
Figure \ref{fig:thermal}(f) shows the read temperature from the cryostats built-in sensor, as well as the estimated temperatures from the resonance and resistance shifts.
Although the sample stage sensor reaches the base temperature within \SI{15}{minutes}, both the resonance position and the resistance take significantly longer to settle to a stable equilibrium. 
Therefore we allow the chip to thermalize before starting a measurement in all subsequent experiments.
The qualitative agreement of the resonance position and resistance also indicates that either method could be used for real-time on-chip monitoring of relative temperature changes and thermal stability. 
The slight overshoot of the estimated temperature from the resonance shift, compared to the estimated temperature from resistance change, can be explained by discrepancies between the used refractive index model and the real index at low temperatures.
The micro-resistor method is particularly well suited for probing the integrated TFLN devices, as it only requires two DC electrical lines and no optical feedthroughs.

\begin{figure*}
    \centering
        \includegraphics[width=1.0\linewidth]{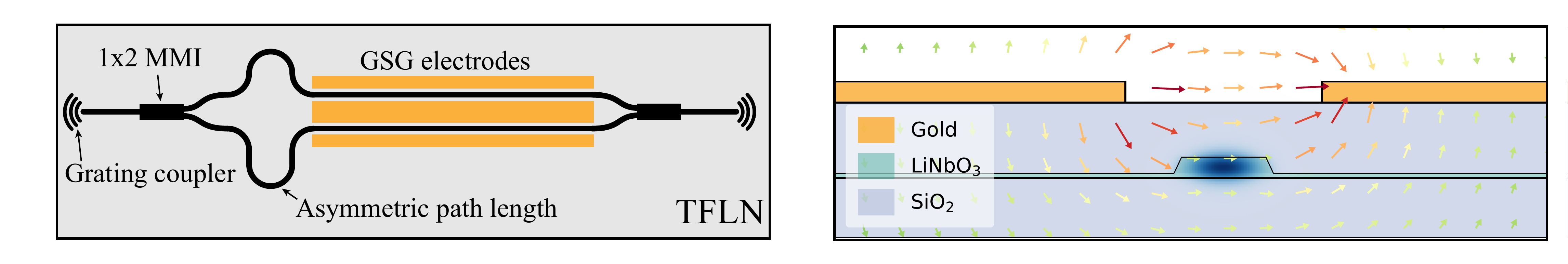}
        \includegraphics[width=1.0\linewidth]{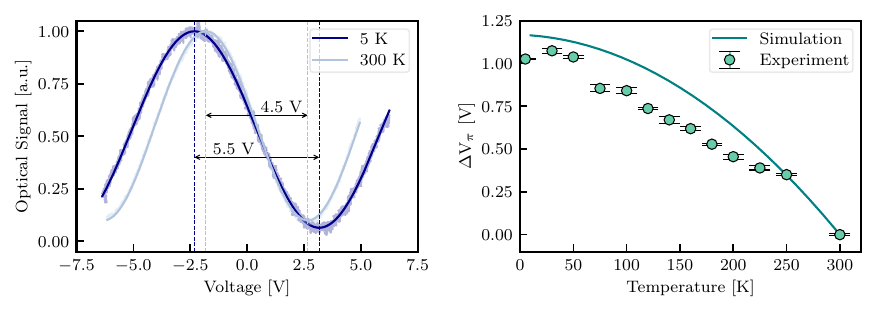}

    \begin{picture}(0,0)
        \put(-230,240){\textbf{a)}}
        \put(0,240){\textbf{b)}}
        \put(-230,160){\textbf{c)}}
        \put(0,160){\textbf{d)}}
    \end{picture}
    \caption{\textbf{TFLN electro-optic modulator.} 
    (a) Device layout used to characterize the $V_\pi$ of a TFLN modulator at cryogenic temperatures.
    (b) Cross-section of the electric field inside a MZM with the magnitude corresponding to the arrow length, and the optical mode in the waveguide.
    (c) Optical response to an applied voltage of a MZM with \SI{6.5}{\mm} long electrodes measured at \SI{5}{K} and \SI{300}{K}.
    (d) The measured $\Delta V_\pi$ for temperatures from \SI{5}{\K} to \SI{300}{\K} (points) and the theoretical curve from simulations (line).
    }
    \label{fig:eo_with_vpi}
\end{figure*}
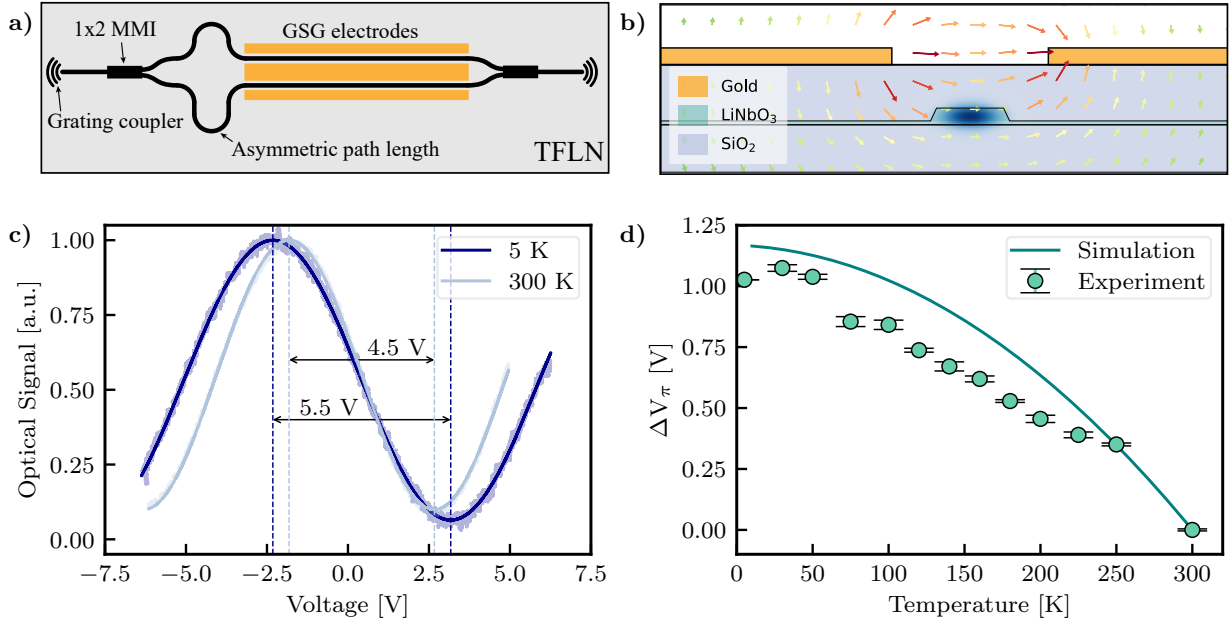

\subsection*{Electro-Optic Modulation}

The electro-optic Pockels effect is one of the most appealing aspects of TFLN photonics, enabling high-speed and low-energy-dissipation optical modulation \cite{hu_integrated_2025}.
We characterize the voltage required for a $\pi$ phase shift ($V_\pi$) of a TFLN Mach-Zehnder modulator (MZM) at different temperatures.
The MZM uses a multimode interferometer (MMI) to split the optical signal into two waveguides and \SI{6.5}{\mm} long gold electrodes, oriented in a push-pull configuration, as shown in the schematic in Fig. \ref{fig:eo_with_vpi}(a).
When applying a triangular wave with a waveform generator, the generated electrical field overlaps with the optical mode in the waveguide, as seen in Fig. \ref{fig:eo_with_vpi}(b), and an index change will be induced in the optical mode due to the electro-optic effect, causing a phase-shift between the two arms of the modulator, and thereby yielding a sinusoidal modulation of the optical signal \cite{wang_integrated_2018}.
In the cryostat, the electrical signal is applied to the TFLN chip using a Ground-Signal-Ground (GSG) RF probe, and the optical signal is coupled via grating couplers.
Figure \ref{fig:eo_with_vpi}(c) shows the sinusoidal response for a TFLN modulator with an electrode separation of $g=\SI{2.8}{\um}$ operating at \SI{10}{kHz}.
At room temperature we measure $V_\pi=\SI{4.5}{V}$ and at \SI{5}{\K} we measure $V_\pi=\SI{5.5}{V}$, corresponding to an increase of \SI{22}{\%} of the required voltage to achieve a $\pi$ phase shift at cryogenic temperatures.
The measured $V_\pi$ at room temperature indicates a $V_\pi\cdot L\approx\SI{2.9}{\V\cdot\cm}$ which is slightly higher than previously reported TFLN modulators with the electrodes deposited on the remaining LN film, compared to our electrodes deposited on top of a \SI{1}{\um} thick SiO$_2$ cladding layer, due to a lower overlap between the electric field and the optical mode \cite{chen_compact_2023}.
To simulate $V_\pi$ of our TFLN modulator, we use the temperature-dependent Pockels coefficient $r_{33}$, the permittivity $\epsilon_{33}$ of lithium niobate as reported in \cite{herzog_electro-optic_2008}, and the varying refractive index given by the Sellmeier equation.
Although these values are reported for a wavelength of \SI{632.8}{\nm} in a temperature range of \SI{7}{\K} to \SI{300}{\K}, we assume that the trend will be equivalent for \SI{1550}{\nm} wavelength.
For each temperature, we simulate the overlap between the electric field and the optical mode $\Gamma$ as well as $r_{33}$ to calculate $V_\pi\cdot L$ as \cite{li_high-performance_2023}
\begin{equation}
    V_\pi\cdot L=\frac{n_{\rm eff}\lambda_0 g}{2n_e^4r_{33}\Gamma},
    \label{eq:vpiL}
\end{equation}
where $n_{\rm eff}$ is the effective mode index, $\lambda_0$ is the vacuum wavelength, and $n_e$ is the LN extraordinary refractive index.
These simulations match the trend of the experiment well, as seen from the $\Delta \rm V_{\pi}$ plotted in Fig. \ref{fig:eo_with_vpi}(d), however, with a minor offset in the absolute value of the $V_{\pi}$, which could be explained by differences in the material properties of the simulations, such as the values for $r_{33}$ and permittivity, or variations in the device geometry.

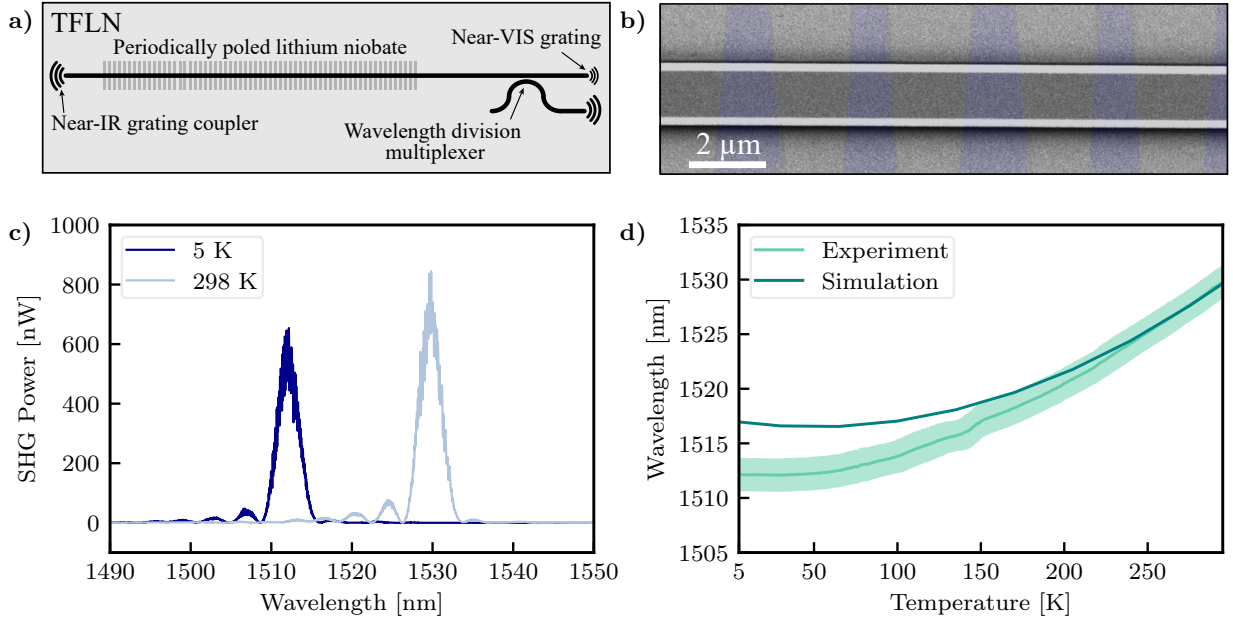
\begin{figure*}
    \centering
        \includegraphics[width=1.0\linewidth]{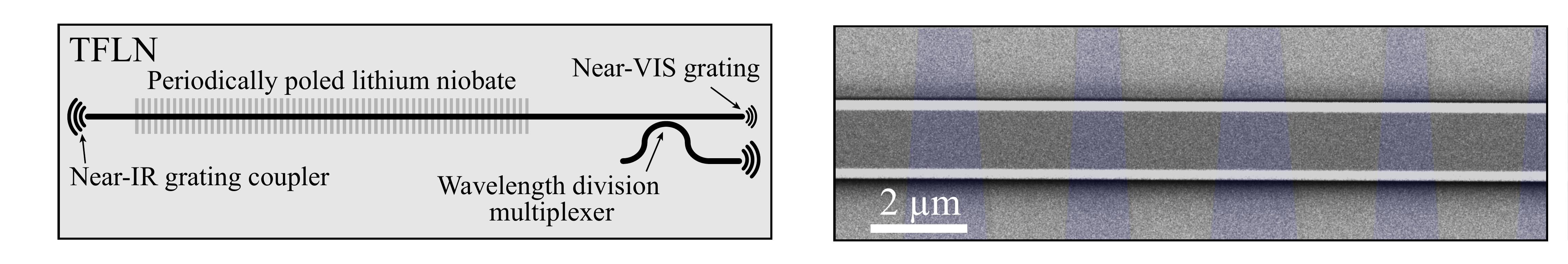}
        \includegraphics[width=1.0\linewidth]{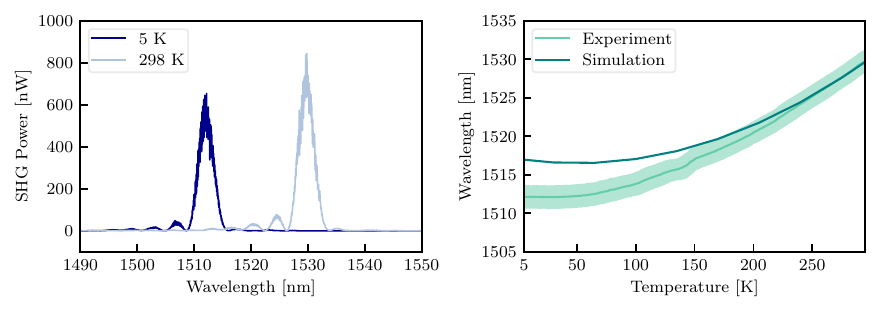}
    \begin{picture}(0,0)
        \put(-230,240){\textbf{a)}}
        \put(0,240){\textbf{b)}}
        \put(-230,160){\textbf{c)}}
        \put(0,160){\textbf{d)}}
    \end{picture}
    \caption{\textbf{Periodically poled TFLN for Type-0 second harmonic generation.} 
    (a) Device layout for measuring second harmonic generation in a periodically poled TFLN waveguide.
    (b) Scanning electron microscopy image of a periodically poled waveguide, with the inverted domains visible in false colour.
    (c) The second harmonic spectrum at \SI{5}{\K} and \SI{298}{\K}.
    (d) Experimental and simulated shift of the central wavelength with respect to temperature. The solid dark line is the simulation from the temperature dependent Sellmeier equations. The light band around the experimental curve indicates the full-width-half-maximum of the measured second harmonic generation spectra.    
    }
    \label{fig:shg_t0}
\end{figure*}

\begin{figure*}
    \centering
        \includegraphics[width=1.0\linewidth]{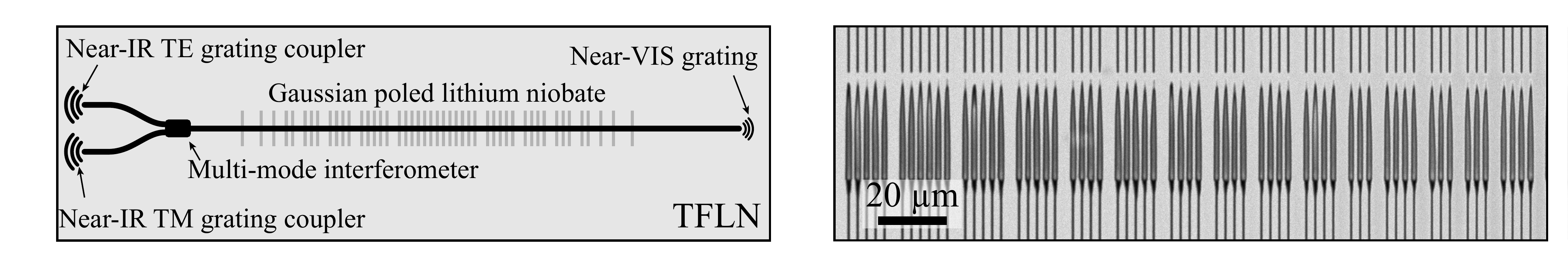}
        \includegraphics[width=1.0\linewidth]{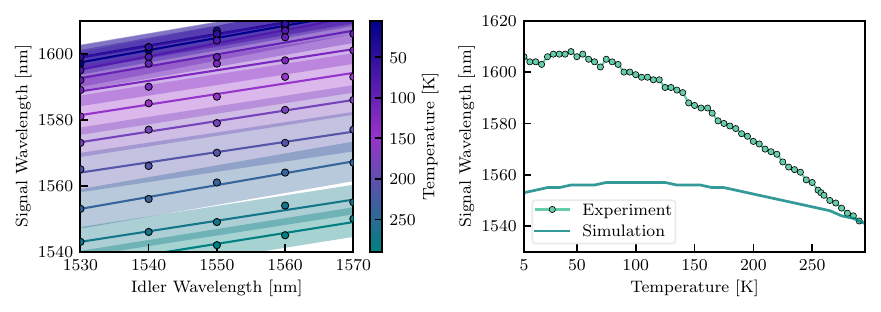}

    \begin{picture}(0,0)
        \put(-230,240){\textbf{a)}}
        \put(0,240){\textbf{b)}}
        \put(-230,160){\textbf{c)}}
        \put(0,160){\textbf{d)}}
    \end{picture}
    \caption{\textbf{Gaussian poled TFLN for Type-II sum frequency generation.} 
    (a) Device layout for measuring Type-II SFG in a Gaussian poled waveguide. 
    (b) Multi-photon microscopy image of the Gaussian modulated domain inversion after the poling process.
    (c) Maxima (line) and full-width-half-maximum (band) of the phase-matching function at different temperatures.
    (d) Measured, as dots, and simulated, as solid line, phase-matched signal wavelength with respect to the sample temperature at fixed idler wavelengths.
    }
    \label{fig:sfg_t2}
\end{figure*}

\subsection*{Type-0 Second Harmonic Generation}

% Device C10 P2 W1.25
Three-wave mixing in TFLN photonics enables another widely used and highly attractive process, nonlinear frequency conversion.
For efficient phase-matching, it is necessary to periodically invert the non-centrosymmetric crystal with the correct period to compensate for the dispersion of the involved optical modes.
Here, we use the \SI{300}{\nm} TFLN platform for Type-0 second harmonic generation (SHG), where all three fields are TE polarized with the crystal axis, which gives the highest efficiency frequency conversion \cite{wang_ultrahigh-efficiency_2018}.
We fabricated a \SI{1.75}{\mm} long periodically poled waveguide with a wavelength division multiplexer (WDM), in the form of a directional coupler, to separate the input light around \SI{1550}{\nm} from the second harmonic light around \SI{775}{\nm}, as seen in Fig. \ref{fig:shg_t0}(a).

To measure the nonlinear response of the waveguide, we sweep the wavelength of a continuous wave laser and use both on-chip and off-chip WDMs to isolate the SHG light, as seen in the schematic in Fig. \ref{fig:shg_t0}(a), with the domains visible in the scanning electron microscopy image in Fig. \ref{fig:shg_t0}(b).
We sweep the laser from \SI{1490}{\nm} to \SI{1550}{\nm}, and tune the chip temperature from \SI{5}{\K} to room temperature.
Figure \ref{fig:shg_t0}(c) shows the SHG response at room temperature and \SI{5}{\K}, exhibiting the characteristic sinc shape of periodic poling quasi-phase-matching, with a shift in the central wavelength of the SHG spectrum by \SI{18}{\nm}, from \SI{1530}{\nm} at room temperature to \SI{1512}{\nm} at \SI{5}{\K}.
Furthermore, Fig. \ref{fig:shg_t0}(d) shows the central phase-matching wavelength and full-width-half-maximum for the full temperature sweep, together with the simulated phase-matching, according to the temperature-dependent Sellmeier equations \cite{edwards_temperature-dependent_1984, jundt_temperature-dependent_1997}, which seems to be accurate above \SI{250}{\K}, but diverges from the experimental data below this temperature.

\subsection*{Type-II Sum Frequency Generation}

While Type-0 phase-matching leverages the highest nonlinear tensor element, Type-II phase-matching makes use of the  $d_{31}$ element, which allows for phase-matching between signal and idler modes of orthogonal polarization.
This unlocks an additional degree of freedom in the dispersion engineering of the TFLN waveguides, which can enable the generation of spectrally uncorrelated photon-pairs, a crucial resource for scalable quantum interference, especially when paired with Gaussian nonlinearity shaping to suppress sidelobes in the phase-matching \cite{kuttner_heralded_2026, xin_spectrally_2022}.
As Type-II phase-matching involves modes of both polarizations, we choose a \SI{600}{\nm} thick TFLN sample with an etch depth of around \SI{400}{\nm}, to support both TE and TM modes at telecom wavelengths, and a waveguide width of \SI{0.8}{\um}, the layout of the used devices can be seen in Fig. \ref{fig:sfg_t2}(a), with a multiphoton microscopy image in Fig. \ref{fig:sfg_t2}(b) showcasing the Gaussian modulated domain inversion after the poling.
We perform sum frequency generation (SFG) with two tunable wavelength lasers in the telecom range, allowing us to map the phase-matching function (PMF) of the \SI{6.5}{\mm} long periodically poled waveguide.
The Type-II process involves signal and idler modes of orthogonal polarization, which is achieved by rotating the polarization of one of the telecom lasers and coupling them both via a polarization maintaining fiber array inside the cryostat.
On-chip, the optical fields of TE and TM polarization are mixed at an MMI and then routed to the poled waveguide.
The generated near-visible signal is TM polarized and out-coupled using a grating coupler, any remaining telecom signal is removed by a fiber WDM.

In Figure \ref{fig:sfg_t2}(c) we plot the maximum of the PMF and its full-width-half-maximum between \SI{5}{\K} and \SI{290}{\K}.
In contrast to the anti-diagonal PMF of the Type-0 processes, the PMF for Type-II is almost diagonal, which is made possible by engineering the waveguide dispersion, thus enabling high-purity heralded photon generation \cite{xin_spectrally_2022, kuttner_heralded_2026}.
It is evident that as the temperature decreases, the PMF shifts from shorter to longer wavelengths on the signal axis.
Figure \ref{fig:sfg_t2}(d) compares the measured and simulated shift of the peak of the signal projection of the PMF with temperature, assuming a fixed idler wavelength.
While the simulation reproduces the correct trend, the predicted shift is considerably smaller in magnitude than the measured one.
We attribute this discrepancy to inaccuracies in the refractive index data at low temperatures, as well as to thermal contraction of the waveguide geometry and poling period.
Accounting for these geometric changes brings the simulation closer to the measured phase-matching, although a significant offset remains.
We can see that with a constant idler wavelength, the required signal wavelength to generate SFG decreases with increasing temperature.
At room temperature with the idler at \SI{1550}{\nano\meter} wavelength, the signal tone phase matches at \SI{1542}{\nano\meter}, whereas at \SI{5}{\K}, the signal phase matches at \SI{1606}{\nano\meter}, indicating a red-shift of \SI{64}{\nano\meter} of the PMF along the signal axis, a trend consistent with experiments in titanium diffused lithium niobate photonics \cite{bartnick_cryogenic_2021, lange_cryogenic_2022}.

\subsection*{Discussion}

In our work, we have studied the optical properties of TFLN photonics in a cryogenic environment, including shifts in optical resonators, electro-optic modulators, and nonlinear frequency converters.
We have first investigated how the thermalization of the TFLN chip can be monitored more precisely, especially during operating temperature changes, with an on-chip micro-resistor compared to the temperature sensor on the sample mount, by comparing the sensor and resistor to the optical response of a racetrack resonator during a cooldown.
The resonance shift and resistor stabilize at the same time, while the sample mount sensor reaches base temperature more rapidly.
We have demonstrated the feasibility and limitations of operating electro-optic modulators in a temperature range from room temperature down to \SI{5}{\K}.
It should also be noted that electro-optic modulators operating at low temperature exhibit reduced slow-speed relaxation, often termed DC drift, which could be beneficial for several quantum information applications that require cryogenic environments \cite{warner_dc-stable_2025}.
The observed increase in modulation voltage ($V_\pi$) of $\sim$\SI{22}{\percent} at \SI{5}{\K} is similar to previous experiments in TFLN \cite{lomonte_single-photon_2021}, however, it is lower than the reported $>$\SI{70}{\percent} increase for titanium in-diffused LN modulators \cite{thiele_cryogenic_2022}.
Our results are consistent with the $\sim$\SI{27}{\percent} decrease of the $r_{33}$ electro-optic coefficient of LN at cryogenic temperatures \cite{herzog_electro-optic_2008}.
The large difference between our measurements with TFLN and previously reported titanium in-diffused waveguides could be due to changes in mode confinement and thermal contraction at low temperatures.
We have investigated the behavior of nonlinear frequency conversion processes at cryogenic temperatures. 
For Type-0 SHG, we have observed a clear blue-shift of \SI{18}{\nm} in the phase-matching at low temperature. 
In contrast, for Type-II SFG we found a red-shift of \SI{64}{\nm} in the phase-matching function along the signal axis, with decreasing temperature.
Although we have reported SFG for non-degenerate wavelengths we can predict the trend of degenerate SHG to longer wavelengths at lower temperatures, which is consistent with bulk lithium niobate observations \cite{bartnick_cryogenic_2021, lange_cryogenic_2022}.
These shifts in the phase-matching wavelengths can be explained by the change of refractive index and dispersion, which is predicted by the temperature-dependent Sellmeier equations \cite{nelson_refractive_1974, smith_refractive_1976, edwards_temperature-dependent_1984, jundt_temperature-dependent_1997}.
While the simulations based on these temperature-dependent Sellmeier equations indicate a trend consistent with the measurement data, for both Type-0 and Type-II phase-matching, the magnitude of the predicted shift is underestimated, especially at lower temperatures.
We attribute this to the coefficients of the Sellmeier equations being reported at and above room temperature, thereby not considering any special change of the refractive index at temperatures closer to absolute zero, which is why they fail to give accurate predictions below \SI{250}{\K} \cite{bartnick_cryogenic_2021}.
It is therefore important to characterize the refractive index and dispersion of LN down to cryogenic temperatures, to give a more robust simulation framework for the prediction of frequency conversion processes.
Overall, the demonstrated shifts in nonlinear frequency conversion processes between room temperature and cryogenic temperatures should be considered when integrated devices are used for low-temperature operation, as the shift in phase-matching can be pre-compensated by biasing the poling period accordingly.

\subsection*{Methods}

\subsubsection*{Nanofabrication}

We use \SI{300}{nm} and \SI{600}{nm} x-cut lithium niobate thin-films on a \SI{2}{\um} thermal oxide insulation layer on a silicon handle.
We start by fabricating pairs of comb-shaped Cr poling electrodes on the surface of the LN film via electron-beam lithography and lift-off.
To invert the orientation of the LN crystals, we apply high-voltage electrical pulses and analyze the results with a commercial multi-photon confocal microscope.
After removing the poling electrodes, we pattern an HSQ mask with an electron-beam and transfer the patterns onto the thin-film using a physical ICP argon plasma etch \cite{kaufmann_redeposition-free_2023}.
Following this, we remove any redeposition of the etching process with a chemical cleaning step, and finally remove any remaining mask by dipping the sample in buffered HF.
This leaves us with waveguides featuring smooth sidewalls with an angle of approximately \SI{65}{\degree}, and a height of \SI{230}{nm} and \SI{400}{nm} respectively.
In order to reduce losses and enhance the device performance, all samples are annealed at \SI{500}{\celsius} for 2 hours, before depositing any electrodes.
For the \SI{300}{nm} film, we deposit \SI{1}{\um} silicon dioxide as a cladding layer, and fabricate gold electro-optic phase shifters on the surface by another round of electron-beam lithography and lift-off.

\subsubsection*{Experimental Setup}

We perform all the swept wavelength experiments with a tunable telecom laser (Toptica CTL 1500) and an InGaAs photodiode for telecom measurements or a Si photodiode for near-VIS measurements. For the SFG sweep, we additionally use a second tunable telecom laser (EXFO TDS200), where we rotate its polarization with a fiber-coupled polarizing beam splitter (OZ Optics).
The chips are mounted in a cryogenic fiber probe station (attoDRY 800), which includes 3 piezo-actuated xyz-stacks (Attocube), as well as an RF GSG probe on a z-stage.
On these stages, there are SMF-28 single-mode fibers mounted, which are rotated to the optimal angle for the coupling via grating couplers. For the SFG experiments, we couple the input signals via a fiber array of PM fibers spaced by \SI{127}{\um}. 

\subsection*{Data availability}
The raw data presented in this study is available from the authors upon reasonable request.

\subsection*{Author Contributions}

US, AM, JK and RJC designed and characterized the 300nm TFLN samples.
AS and GF fabricated the 300nm TFLN samples.
TK designed the 600nm TFLN samples.
TK and AS fabricated the 600nm TFLN samples.
TK and MR characterized the 600nm TFLN samples.
US, AM, JK, TK and RJC performed data analysis.
TK and RJC wrote the manuscript.
All authors contributed to revising the manuscript.
AM, RJC and RG supervised the project.

\subsection*{Acknowledgments}

We acknowledge support from the Scientific Center of Optical and Electron Microscopy ScopeM and from the cleanroom facilities BRNC and FIRST of ETH Zurich.
R.J.C. and T.K. acknowledge support from the Swiss National Science Foundation under the Ambizione Fellowship Program (Project Number 208707) and support from an ETH Research Grant.
R.G. acknowledges support from the European Space Agency (Project Number 4000137426), the Swiss National Science Foundation under the Bridge Program (Project Number 194693) and the Sinergia Program (Project Number 206008), and the European Research Council (Project Number 714837).

%\bibliography{references}

%apsrev4-2.bst 2019-01-14 (MD) hand-edited version of apsrev4-1.bst
%Control: key (0)
%Control: author (8) initials jnrlst
%Control: editor formatted (1) identically to author
%Control: production of article title (0) allowed
%Control: page (0) single
%Control: year (1) truncated
%Control: production of eprint (0) enabled
%

\end{document}